\documentclass[12pt]{article}

\input{epsf}
\usepackage[dvips]{graphicx}
\usepackage{cite}
\def\1ad{\mbox{\normalsize $^1$}}
\def\2ad{\mbox{\normalsize $^2$}}
\def\3ad{\mbox{\normalsize $^3$}}
\def\4ad{\mbox{\normalsize $^4$}}

\def\5ad{\mbox{\normalsize $^5$}}
\def\6ad{\mbox{\normalsize $^6$}}
\def\7ad{\mbox{\normalsize $^7$}}
\def\8ad{\mbox{\normalsize $^8$}}

\def\GE73{{}^{73}{\rm Ge}}
\def\gE76{{}^{76}{\rm Ge}}
\def\gEe74{{}^{74}{\rm Ge}}
\def\xe131{{}^{131}{\rm Xe}}
\def\i127{{}^{127}{\rm I}}

\def\dj{\hbox{d\kern-0.347em \vrule width 0.3em height 1.252ex depth
-1.21ex \kern 0.051em}}

\newcommand{\be}{\begin{equation}}
\newcommand{\ee}{\end{equation}}
\newcommand{\ben}{\begin{equation*}}
\newcommand{\een}{\end{equation*}}
\newcommand{\ba}{\begin{eqnarray}}
\newcommand{\ea}{\end{eqnarray}}
\newcommand{\ban}{\begin{eqnarray*}}
\newcommand{\ean}{\end{eqnarray*}}
\newcommand{\brr}{\begin{array}}
\newcommand{\err}{\end{array}}
\newcommand{\bc}{\begin{center}}
\newcommand{\ec}{\end{center}}

\newcommand{\bea}{\begin{eqnarray}}
\newcommand{\eea}{\end{eqnarray}}
\newcommand{\bean}{\begin{eqnarray*}}
\newcommand{\eean}{\end{eqnarray*}}

\newcommand\lsim{\mathrel{\rlap{\lower4pt\hbox{\hskip1pt$\sim$}}
    \raise1pt\hbox{$<$}}}
\newcommand\gsim{\mathrel{\rlap{\lower4pt\hbox{\hskip1pt$\sim$}}
    \raise1pt\hbox{$>$}}}

\newcommand{\centeron}[2]{{\setbox0=\hbox{#1}\setbox1=\hbox{#2}\ifdim
                             \wd1>\wd0\kern.5\wd1\kern-.5\wd0\fi \copy0
                             \kern-.5\wd0\kern-.5\wd1\copy1\ifdim\wd0>\wd1
                             \kern.5\wd0\kern-.5\wd1\fi}}
\newcommand{\ltap}{\>\centeron{\raise.35ex\hbox{$<$}}
                     {\lower.65ex\hbox{$\sim$}}\>}
\newcommand{\gtap}{\>\centeron{\raise.35ex\hbox{$>$}}
                     {\lower.65ex\hbox{$\sim$}}\>}

\begin{document} 

\setcounter{page}{0}
\thispagestyle{empty}

\begin{flushright}
SPhT-T05/019\\
hep-ph/yymmxxx
\end{flushright}

\vskip 8pt

\begin{center}
{\bf \Large {
Indirect Detection of \\ [0.25cm]  
Dirac Right-Handed Neutrino Dark Matter 
}}
\end{center}

\vskip 10pt

\begin{center}
{\large Dan Hooper $^{a}$ and G\'eraldine Servant $^{b}$ }
\end{center}

\vskip 20pt

\begin{center}

\centerline{$^{a}$ {\it Astrophysics, University of Oxford, Oxford, UK}}
\vskip 3pt
\centerline{$^{b}${\it Service de Physique Th\'eorique, CEA Saclay, F91191 Gif--sur--Yvette,
France}}
\vskip .3cm
\centerline{\tt  hooper@astro.ox.ac.uk, servant@spht.saclay.cea.fr}
\end{center}

\vskip 13pt

\begin{abstract}
\vskip 3pt
\noindent

We present the signatures and prospects for the indirect detection of a Dirac right-handed neutrino dark matter candidate in neutrino telescopes, cosmic positron experiments and gamma-ray telescopes. An example of such a dark matter candidate can be found in extra-dimensional models. In some constructions, Kaluza--Klein states with the gauge quantum numbers of a right-handed neutrino can have sizable gauge interactions with Standard Model particles. For instance, in 5D warped Grand Unified Theories, it has been shown that
a Kaluza--Klein right-handed neutrino may be stable and otherwise a phenomenologically viable dark matter candidate. We find that the prospects for the indirect detection of such a WIMP are encouraging, particularly for neutrino telescopes and cosmic positron experiments.

\end{abstract}

\vskip 13pt
\newpage
\section{Introduction}

Despite our ignorance of the nature of dark matter, numerous techniques have been developed in an effort to detect it \cite{review}. Indeed, if the dark matter consists of Weakly Interacting Massive Particles (WIMPs), as is suggested by our theoretical prejudices, it will have a tiny but non-vanishing probability of interacting with the ordinary matter of the Standard Model (SM).

In addition to collider searches, direct and indirect detection techniques have been devised to search for particle dark matter. In the case of direct detection, one attempts to measure the recoil energy of target nuclei in an underground detector due to its elastic scattering with a WIMP. Thousands of WIMPs are expected to cross each square centimeter every second at the surface of the Earth. We could potentially detect up to a few tens of events per year at near future detectors, depending on the precise interactions of the WIMP.

The goal of indirect detection techniques on the other hand, is to detect a flux of cosmic rays resulting from the annihilation of WIMPs in regions such as the Galactic center, the Galactic halo, or the interior of the Sun. These annihilations can produce potentially observable fluxes of gamma-rays, neutrinos, positrons, anti-protons and anti-deuterons. 



Given the abundance of experimental activity related to dark matter detection, it is timely to study the distinctive signatures expected in different dark matter scenarios. Such signatures and event rates can vary substantially from one WIMP model to another. The most studied candidate so far is the Lightest Supersymmetric Particle (LSP) in models of R-parity conserving supersymmetry. Some alternatives to neutralinos and other LSPs do exist, however. Lately, the possibility of Kaluza--Klein dark matter has been explored. While the idea that dark matter could be made of Kaluza--Klein (KK) particles is very tempting, it turns out that this is not so easy to achieve. Indeed, in most extra-dimensional models, there are no stable KK states, all being able to decay to SM particles. An exception is the class of models in Universal Extra Dimensions (UED) \cite{Appelquist:2000nn}. In this case, all SM fields propagate in flat toroidal extra dimensions. Translation
invariance along an extra dimension is only broken by the orbifold imposed to recover a chiral SM spectrum. Still, there is a remnant discrete symmetry called KK parity, $(-1)^n$, where $n$ is the KK number. This symmetry insures that interaction vertices cannot involve an odd number of odd-KK states and, therefore, a vertex with two SM particles (with $n=0$) and one KK state (with $n=1$) is forbidden. As a result, the Lightest KK Particle (LKP) with $n=1$ cannot decay into SM particles and is stable. For $\sim$TeV$^{-1}$ sized extra dimensions, the LKP can act as a WIMP. Relic density \cite{Servant:2002aq,Kakizaki:2005en}, 
direct \cite{Cheng:2002ej,Servant:2002hb} and indirect detection \cite{Cheng:2002ej,Hooper:2002gs,Bertone:2002ms,Hooper:2004xn,Bergstrom:2004cy,Baltz:2004ie,Bergstrom:2004nr}  studies of this candidate have all been carried out in the last few years. Constraints on these models from radion cosmology have also been studied \cite{Kolb:2003mm}.

The second example of KK dark matter arises in the context of warped geometries and more specifically in  the context of warped Grand Unified Theories (GUTs) \cite{Agashe:2004ci,Agashe:2004bm}. To locate these constructions in the landscape of extra dimensional models, recall that the interest in the phenomenology of extra dimensions over the last few years has been motivated by the goal of understanding the weak scale. The only extra-dimensional geometry which really addresses the hierarchy problem is the Randall-Sundrum geometry. 
 Particle physics model building in this framework has been flourishing and a favorite class of models has emerged: that where all SM fields propagate in the bulk of AdS$_5$, except for the Higgs (or alternative physics responsible for electroweak symmetry breaking) which is localized on the IR brane. In addition, the electroweak gauge group should be extended to $SU(2)_L\times SU(2)_R\times U(1)$. Those models were embedded into a GUT in Refs.~\cite{Agashe:2004ci,Agashe:2004bm} and it is in this context that a viable dark matter appears.

In these models, a stable KK fermion can arise as a consequence of imposing proton stability in a way very reminiscent to R-parity stabilizing the LSP in supersymmetric models. The symmetry is called $Z_3$ and is a linear combination of baryon number and $SU(3)$ color. It actually exists in the SM but SM particles are not charged under it since only colored particles carry baryon number in the SM. In  Refs.~\cite{Agashe:2004ci,Agashe:2004bm}, and more generally in higher dimensional GUTs, baryon number can be assigned in such a way that there exists exotic KK states with the gauge quantum numbers of a lepton and which carry baryon number as well as KK quarks which carry non-standard baryon number. These particles carry a non-zero $Z_3$ charge. The lightest of these, called the Lightest $Z_3$ Particle (LZP), is stable since it cannot decay into SM particles.

The only massive elementary Dirac fermion (with a mass in the 1 GeV - 1 TeV range) which could be a viable dark matter candidate is the neutrino. If such a neutrino had the same coupling to the $Z$ as in the SM, however, it would be excluded by direct detection experiments. Its coupling to the $Z$, therefore, must be suppressed\footnote{Note that in SUSY, such constraints are weaker because of the Majorana nature of the neutralino.}. Thus, we are left with the possibility of a  KK Right-Handed (RH) neutrino. In models where the electroweak gauge group is extended to $SU(2)_L\times SU(2)_R\times U(1)$, the RH neutrino has gauge interactions in particular with the additional $Z^{\prime}$. Nevertheless, its interactions with ordinary matter are feeble because they involve the additional gauge bosons which have a large mass ($\gsim 3$ TeV). This opens the possibility of a weakly interacting Dirac RH neutrino. 

In this work, we present the prospects for the indirect detection of right-handed neutrino dark matter, based on the particular example of the LZP. In section \ref{LZPpresentation}, we will present the properties of this particle, how it gets its mass and how it interacts with ordinary matter.
Section \ref{sec:Sun}  deals with the predictions for the detection of neutrinos generated in annihilation of the LZP in the Sun. Signatures due to cosmic positrons from annihilation in the galactic halo are presented in section \ref{sec:positrons} and gamma-rays from annihilation in the Galactic center in section \ref{sec:GC}.
We summarize our results and make comparisons with LSP and LKP dark matter candidates in the following section.

\section{A model of Dirac right-handed neutrino dark matter}
\label{LZPpresentation}

As mentioned in the introduction, the LZP arises as a viable dark matter candidate in warped GUTs. An interesting feature of these models is that 
GUT states such as $X, Y$ gauge bosons appear at the TeV scale (via the KK excitations). In $SO(10)$, there are also the $X^{\prime}, Y^{\prime}$,  $X_s$, $Z^{\prime}$, etc. that the LZP can couple to.

The LZP is a Kaluza--Klein fermion, which is a four-component spinor and vector-like object. It gets its mass by marrying the left-handed with the right-handed chirality. As explained in great detail in Ref.~\cite{Agashe:2004bm}, it can be naturally very light, much lighter than the KK scale of Randall-Sundrum models, namely $M_{KK}\gsim$ 3 TeV. This is because the RH chirality is localized near the TeV brane while the LH one is near the Planck brane. The overlap of wave functions is small, resulting in a small Dirac KK mass.  Its lightness is related to the top quark's heaviness but not entirely fixed by it, so that we will consider LZPs in the mass range of approximately 20 GeV to a few TeV. In this paper, we refer to the LZP as if it were a chiral fermion 
 because only the RH chirality has significant interactions and the other chirality decouples. In addition, the LZP 
has the same gauge quantum numbers as the RH neutrino of $SO(10)$ or Pati--Salam. As a result, we refer to it as a ``Dirac RH neutrino" and denote it as $\nu^{\prime}_R$.

The LZP couples to the TeV KK gauge bosons of $SO(10)$. In addition, when electroweak symmetry is broken, $Z-Z^{\prime}$ mixing induces a coupling of the RH neutrino to the SM $Z$ gauge boson. This coupling is suppressed by $(M_Z/M_{Z^{\prime}})^2$. If $M_{Z^{\prime}}\sim$ few TeV (the mass of KK gauge bosons is set by $M_{KK} $), the size of this coupling will typically be ideal for a WIMP. There is actually a second source for the $Z$-LZP coupling, which we will not discuss here but refer the reader to Ref.~\cite{Agashe:2004bm} for a detailed explanation. We denote $g_Z^{ \nu^{\prime}_R}$ as the coupling of $\nu_R^{\prime}$ to the Z. There is enough freedom in the model under consideration to treat this coupling as an almost arbitrary parameter. 
 
As far as indirect detection is concerned, three phenomenological quantities are especially important: the annihilation cross section of the LZP, the fraction of annihilations which produce various SM particles and the LZP's elastic scattering cross section with nuclei (for the WIMP's capture in the Sun). These quantities were described in detail in Ref.~\cite{Agashe:2004bm}.
For LZPs lighter than approximately 100 GeV, LZP annihilations proceed dominantly via s-channel $Z$-exchange. For larger masses, annihilation via the t-channel exchange of $X_s$ into top quarks or via s-channel $Z^{\prime}$ exchange into $t \overline{t}$, $b \overline{b}$, $W^+ W^-$ and $Z h$ dominates. 

Concerning elastic scattering,  as is well-known for a Dirac neutrino, the Spin Independent (SI) elastic scattering cross section via a t-channel Z exchange has the form
\begin{equation}
 \sigma_{SI}\propto  \left[Z(1-4\sin^2\theta_{\rm{W}})-(A-Z)\right]^2. 
 \label{sigmaSI0}
 \end{equation}
Since $4\sin^2\theta_{\rm{W}} \approx 1$, the coupling to protons is suppressed. In these models, the Spin-Dependent (SD) elastic scattering cross section can also play an important role, although only through scattering off of hydrogen in the Sun, since the spin of $^4$He is zero.

\begin{figure}[h]
\begin{center}
\includegraphics[width=5.40cm,height=4.5cm]{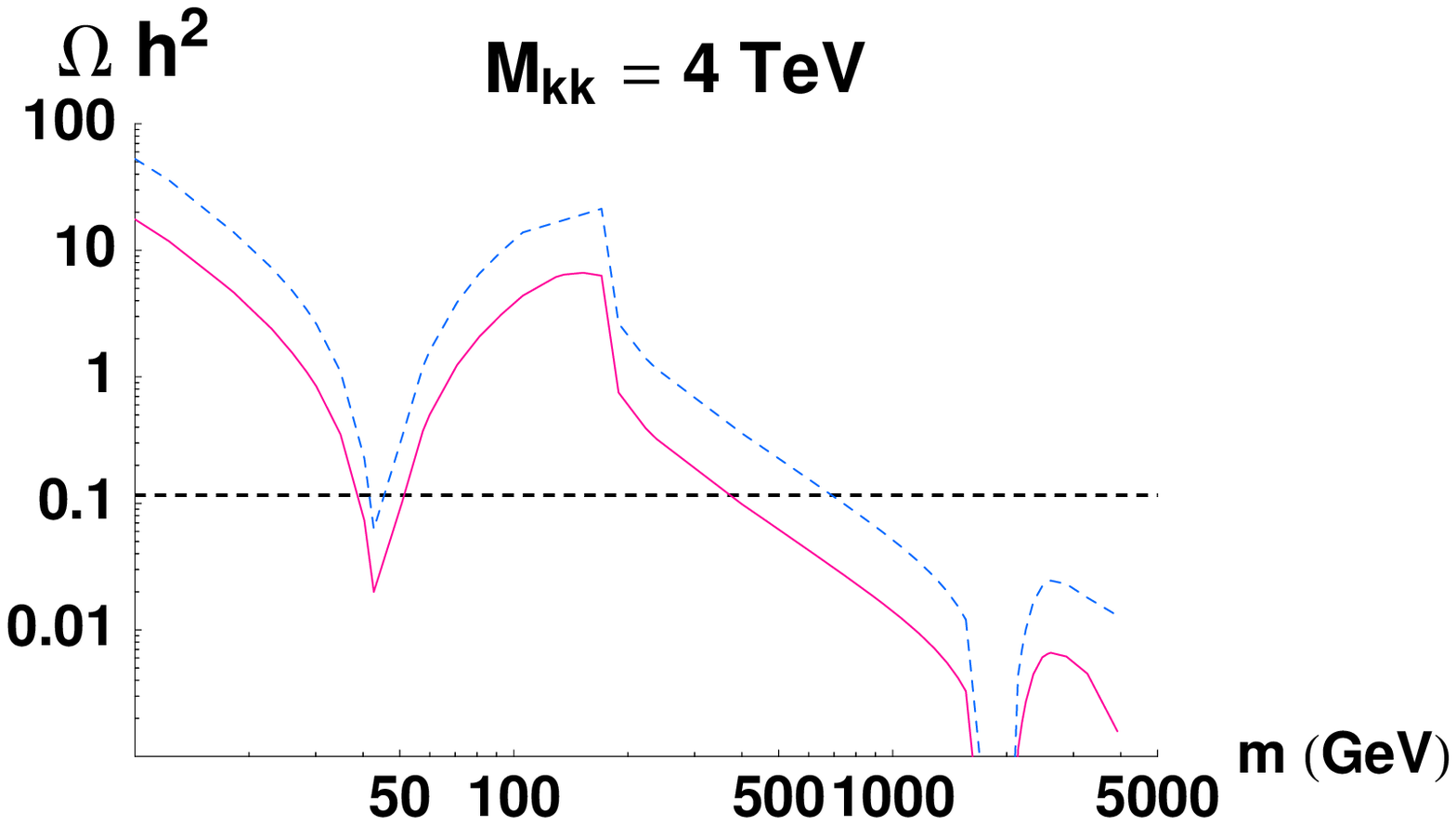}
\includegraphics[width=5.40cm,height=4.5cm]{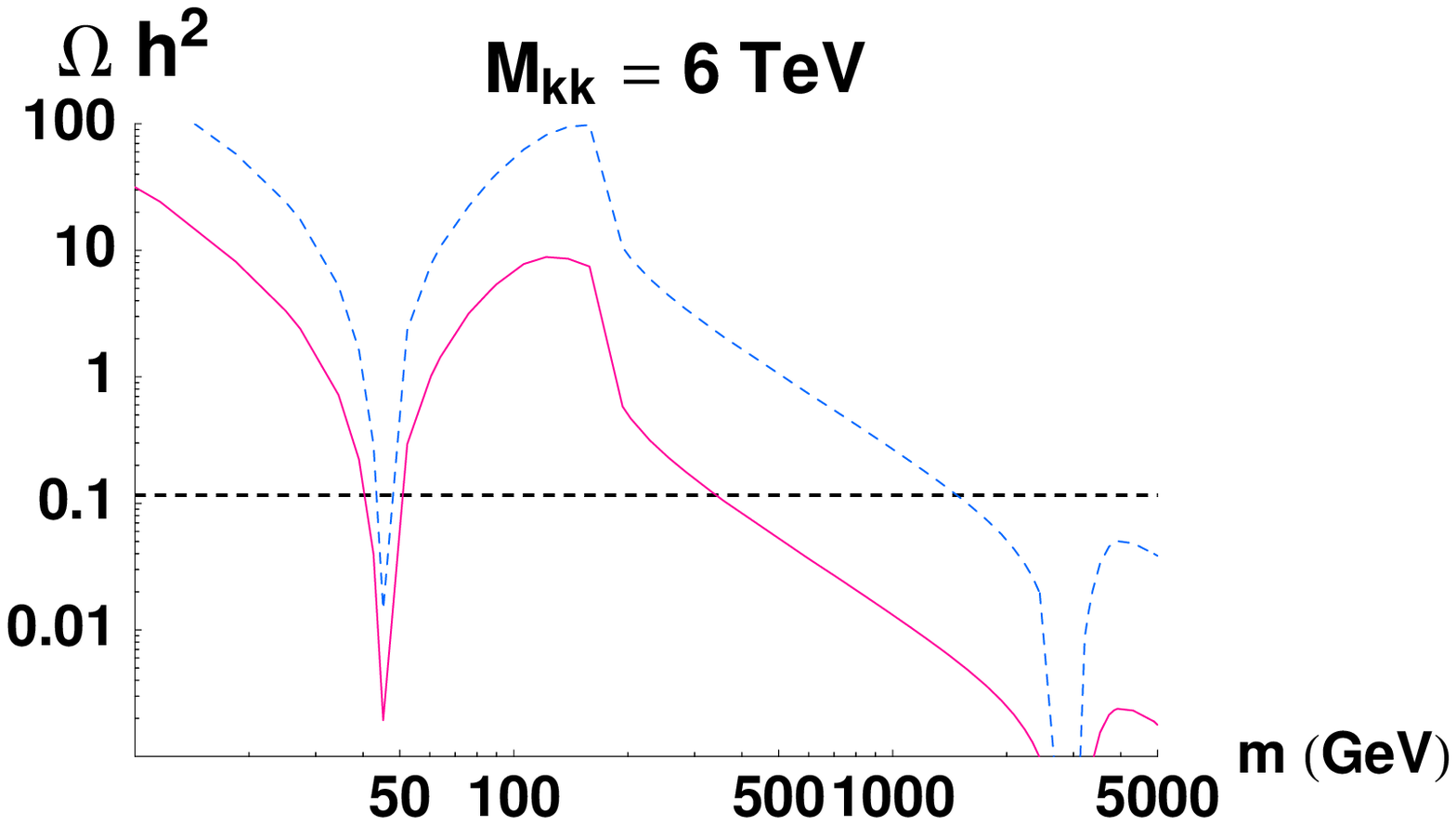}
\includegraphics[width=5.40cm,height=4.5cm]{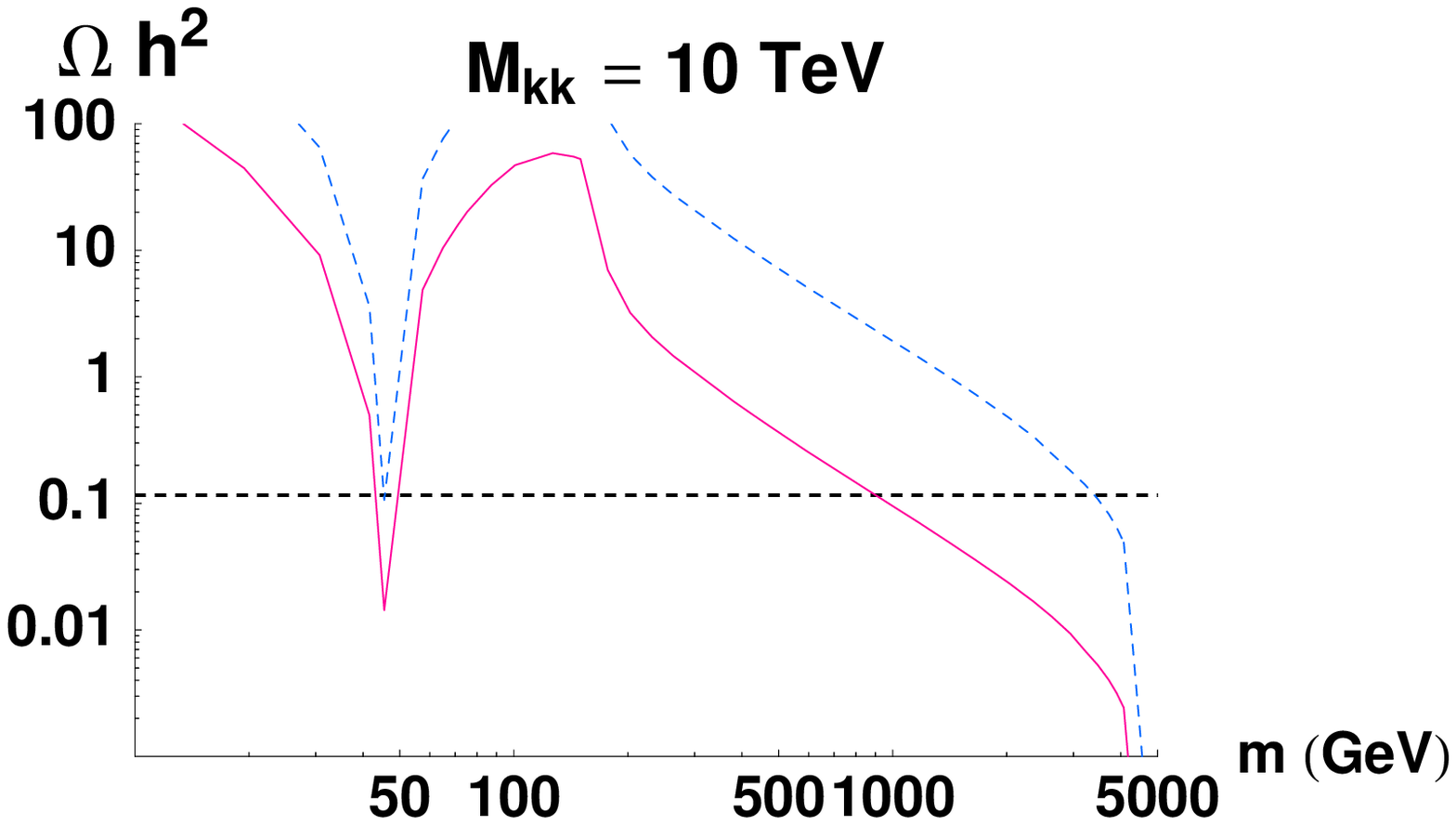}
\caption[]{Predictions for $\Omega_{\mbox{\tiny LZP}} h^2$ as a function of the LZP mass for three values of the KK gauge boson mass, $M_{KK}$. In each case, the upper dashed blue line corresponds to the `minimal' value of the $Z$-LZP coupling, $g_Z^{ \nu^{\prime}_R}$, while the lower pink curve is for a `median' value of this coupling. In this article, we study the regions of parameter space given in the third column of  Table \ref{datasets-table}.}
\label{relic}
\end{center} 
\end{figure}

As presented in Ref.~\cite{Agashe:2004bm}, there is a large region of parameter space in these models. The value of $g_Z^{ \nu^{\prime}_R}$ depends sensitively on where we sit in this parameter space. In this study, we focus only on the regions in which the total annihilation cross section leads to a thermal relic density which does not exceed the measured value. It is indeed conceivable that the relic density we observe today is not only due to the contribution obtained from the standard thermal calculation, but also from non-thermal processes, such as the decay of heavier KK modes, gauge bosons or fermions. We have thus selected three representative regions of parameter space, corresponding to $M_{KK}=4$, 6 and 10 TeV. For each $M_{KK}$ value, we have two sets of values for $g_Z^{ \nu^{\prime}_R}$ named `minimal' and `median'. The corresponding predictions for the dark matter relic density are given in Fig. \ref{relic} and are summarized in Table \ref{datasets-table}. 
%
\begin{table}[h]
\begin{center}
\renewcommand{\arraystretch}{.4}\small\normalsize
\begin{tabular}{rc|l|ll} \hline\hline 
&&& \\
\multicolumn{2}{c}{Data sets} & \multicolumn{1}{|c}{Elastic scattering}  &  \multicolumn{1}{|c}{LZP mass range}\\
& &  cross section, $\sigma_{\mbox{\tiny LZP}-\rm{nucleon}}$  & \hspace{0.3cm}    leading to $\Omega h^2 \lsim 0.11$ \\ \hline  
&&& \\

$M_{KK}=$4 TeV, &Minimal $g_Z^{ \nu^{\prime}_R}$ & $\sim 6 \times 10^{-8}$ pb &40 GeV $ \lsim m_{\mbox{\tiny LZP}} \lsim 45 $ GeV,\\ 
&&& \\
&&& $m_{\mbox{\tiny LZP}} \gsim 700 $ GeV \\ \hline
&&& \\
$M_{KK}=$4 TeV, &Median  $g_Z^{ \nu^{\prime}_R}$ & $\sim 2 \times 10^{-7}$ pb &35 GeV $ \lsim m_{\mbox{\tiny LZP}} \lsim 50 $ GeV,\\ 
&&& \\
&&& $m_{\mbox{\tiny LZP}} \gsim 350 $ GeV \\ \hline \hline
&&& \\
$M_{KK}=$6 TeV, &Minimal $g_Z^{ \nu^{\prime}_R}$ & $\sim  10^{-8}$ pb &41 GeV $ \lsim m_{\mbox{\tiny LZP}} \lsim 48 $ GeV,\\
&&& \\
&&& $m_{\mbox{\tiny LZP}} \gsim 1500 $ GeV \\ \hline
&&& \\

$M_{KK}=$6 TeV, &Median  $g_Z^{ \nu^{\prime}_R}$ & $\sim  10^{-7}$ pb &40 GeV $ \lsim m_{\mbox{\tiny LZP}} \lsim 50 $ GeV,\\ 
&&& \\
&&& $m_{\mbox{\tiny LZP}} \gsim 310 $ GeV \\ \hline \hline
&&& \\

$M_{KK}=$10 TeV, &Minimal $g_Z^{ \nu^{\prime}_R}$ & $\sim  10^{-9}$ pb & $ m_{\mbox{\tiny LZP}} \gsim 3100 $ GeV \\ 
&&& \\ \hline
&&& \\
$M_{KK}=$10 TeV, &Median  $g_Z^{ \nu^{\prime}_R}$ & $\sim  10^{-8}$ pb &40 GeV, $ \lsim m_{\mbox{\tiny LZP}} \lsim 50 $ GeV\\ 
&&& \\
&&& $m_{\mbox{\tiny LZP}} \gsim 900 $ GeV \\ \hline \hline
\end{tabular}
\end{center}
\caption{Data sets used in our analysis (in particular in Fig.\ref{icecubeevents}). They  can lead to the right relic density (see Fig \ref{relic})
 and do not exceed direct detection limits.}
\label{datasets-table}
\end{table}
\section{Neutrinos From LZP Annihilations in the Sun}
\label{sec:Sun}

WIMPs travelling through the Galactic halo will occasionally scatter off of large bodies, such as the Sun, and become gravitationally bound. The capture rate for LZPs in the Sun is given by~\cite{capture}:
\begin{equation}
C^{\odot} \simeq 4.5 \times 10^{18} \, \mathrm{s}^{-1} 
\left( \frac{\rho_{\mathrm{local}}}{0.4\, \mathrm{GeV}/\mathrm{cm}^3} \right) 
\left( \frac{270\, \mathrm{km/s}}{\bar{v}_{\mathrm{local}}} \right)^3  
\left( \frac{\sigma_{\mathrm{H, SD}}+ \sigma_{\mathrm{H, SI}}
+ 0.067 \, \sigma_{\mathrm{He, SI}} } {10^{-6}\, \mathrm{pb}} \right)
\left( \frac{1000 \, \mathrm{GeV}}{m_{\rm{LZP}}} \right)^2, 
\end{equation}
where $\rho_{\mathrm{local}}$ and $\bar{v}_{\mathrm{local}}$ are the local density and rms velocity of halo dark matter particles, respectively. $\sigma_{\mathrm{H,SD}}$, $\sigma_{\mathrm{H, SI}}$ and $\sigma_{\mathrm{He, SI}}$ are the LZP's elastic Spin-Dependent (SD) and Spin-Independent (SI) scattering cross sections with hydrogen and helium nuclei. The cross section for LZP spin-independent scattering off of neutrons is much larger than for protons, making $\sigma_{\mathrm{H, SI}}$ unimportant compared to $\sigma_{\mathrm{H,SD}}$ and $\sigma_{\mathrm{He, SI}}$.

The spin-independent cross sections in this model are \cite{Agashe:2004bm}:
 \begin{equation}
 \sigma_{H,SI}=\frac{ \left( g_Z^{ \nu^{\prime}_R } \right)^2 
e^2 \mu^2_H (1-4\sin^2 \theta_{\rm{W}})^2}{64 \pi M_Z^4 \sin^2 \theta_{\rm{W}} \cos^2 \theta_{\rm{W}}} \ \ \ , \ \ \ 
  \sigma_{He,SI}=\frac{ \left( g_Z^{ \nu^{\prime}_R } \right)^2 
e^2 \mu^2_{He} \left(2(1-4\sin^2 \theta_{\rm{W}})-2\right)^2}{64 \pi M_Z^4 \sin^2 \theta_{\rm{W}} \cos^2 \theta_{\rm{W}}},
 \end{equation}
where $ \mu_{H,He}=m_{\mbox{\tiny LZP}} m_{H,He}/(m_{\mbox{\tiny LZP}}+m_{H,He})$. Note that $\sigma_{He,SI}$ dominates by a factor $10^4$ over  $\sigma_{H,SI}$.

The cross section for the SD interaction is given by \cite{Agashe:2004bm}:
\begin{equation}
 \sigma_{H,SD}= \frac{3 \ \mu^2_{H} \Lambda^2}{\pi} \  , \ \ \  \Lambda=a_p
  \ , \ \ \ 
 a_{p}=\frac{e \ g_Z^{\nu'_R} } {8 M_Z^2 \cos \theta_{\rm{W}} \sin \theta_{\rm{W}}} \left[ -\Delta u + \Delta d+\Delta s\right],
\end{equation}
where
$\Delta u=0.78\pm 0.02$, $\Delta d=-0.48\pm 0.02$ and $\Delta s=-0.15\pm 0.02$ \cite{Ellis:2000ds}
 are the spins carried by the u, d and s quarks in the proton, respectively.

\begin{figure}[h]
\begin{center}
\includegraphics[width=9.0cm,height=4.5cm]{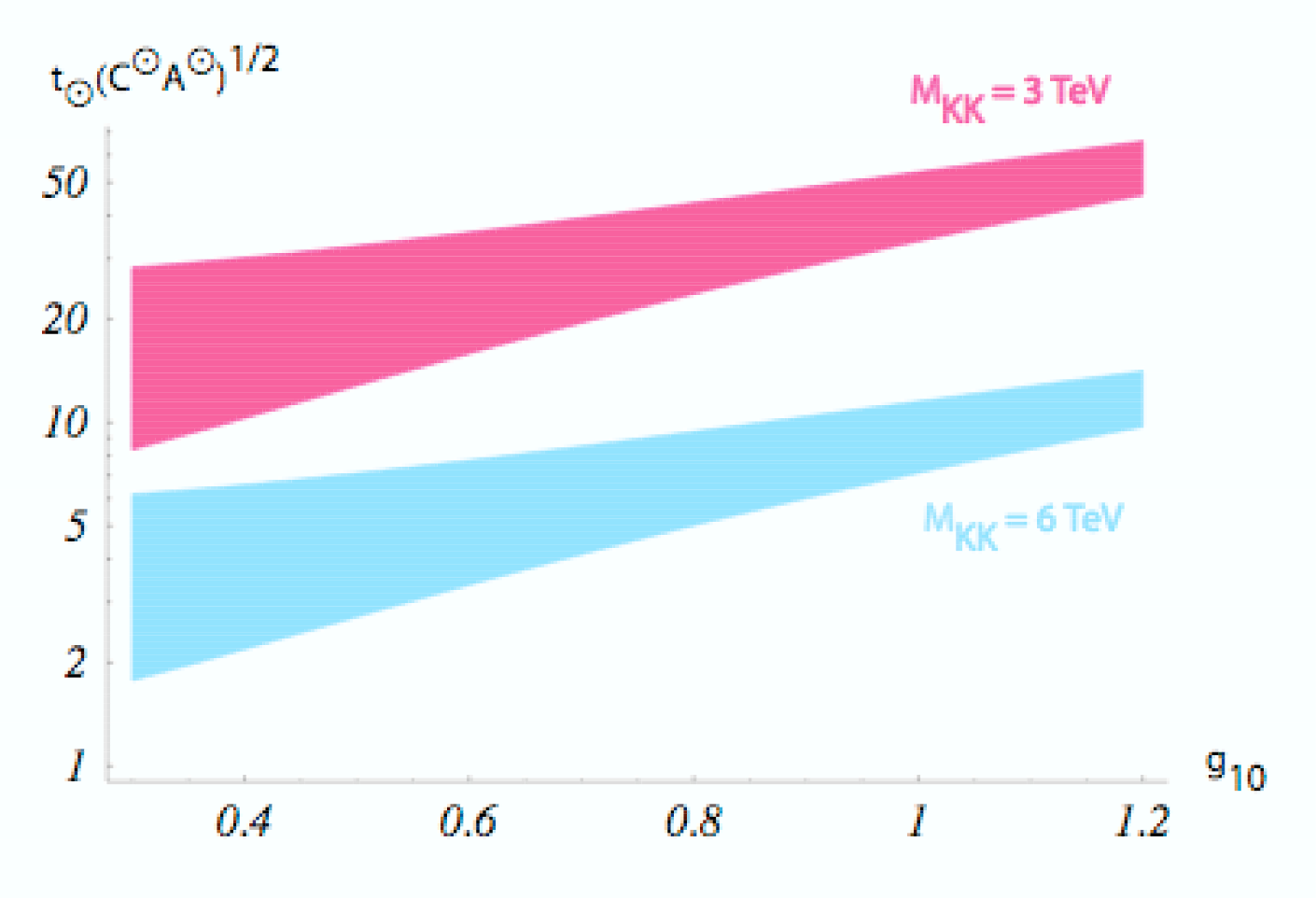}
\caption[]{$\sqrt{C^{\odot} A^{\odot}} t_{\odot}$ evaluated at $\langle \sigma  v \rangle \approx 3 \times 10^{-26}$ cm$^3$/s, the value of the annihilation cross section leading to the correct thermal relic density. Throughout the parameter space, the Sun consistently reaches equilibrium between the LZP capture and annihilation rates. Figure from Ref.~\cite{Agashe:2004bm}.}
\label{equil}
\end{center} 
\end{figure}

The time rate of change of the number of LZPs in the Sun is given by:
\begin{equation}
\dot{N} = C^{\odot} - A^{\odot} N^2 \, ,
\end{equation}
where $C^{\odot}$ is the capture rate and $A^{\odot}$ is the 
annihilation cross section times the relative WIMP velocity per volume. $A^{\odot}$ is given by:
\begin{equation}
A^{\odot} = \frac{\langle \sigma v \rangle}{V_{\mathrm{eff}}}, 
\end{equation}
where $V_{\mathrm{eff}}$ is the effective volume of the core
of the Sun and is determined by matching the core temperature with 
the gravitational potential energy of a single WIMP at the core
radius.  This was found to be~\cite{equ}:
\begin{equation}
V_{\rm eff} = 1.8 \times 10^{26} \, \mathrm{cm}^3 
\left( \frac{1000 \, \mathrm{GeV}}{m_{\rm{LZP}}} \right)^{3/2} \, .
\end{equation}
The present day annihilation rate of LZPs in the Sun is given by:
\begin{equation} 
\Gamma = \frac{1}{2} A^{\odot} N^2 = \frac{1}{2} \, C^{\odot} \, 
\tanh^2 \left( \sqrt{C^{\odot} A^{\odot}} \, t_{\odot} \right) \, , 
\end{equation}
where $t_{\odot} \simeq 4.5$ billion years is the age of the solar system. If the capture rate in the Sun and annihilation cross section are large enough, these two processes will reach equilibrium with each other. This occurs for $ \sqrt{C^{\odot} A^{\odot}} \, t_{\odot} \gsim 1$. In fig.~\ref{equil}, $\sqrt{C^{\odot} A^{\odot}} t_{\odot}$ is evaluated at $\langle \sigma  v \rangle \approx 3 \times 10^{-26}$ cm$^3$/s, the value of the annihilation cross section leading to the correct relic density. Throughout the parameter region leading to the observed relic density, the Sun consistently reaches equilibrium between the LZP capture and annihilation rates.
%

\begin{figure}[t]
\centering\leavevmode
\mbox{
\includegraphics[width=3.2in]{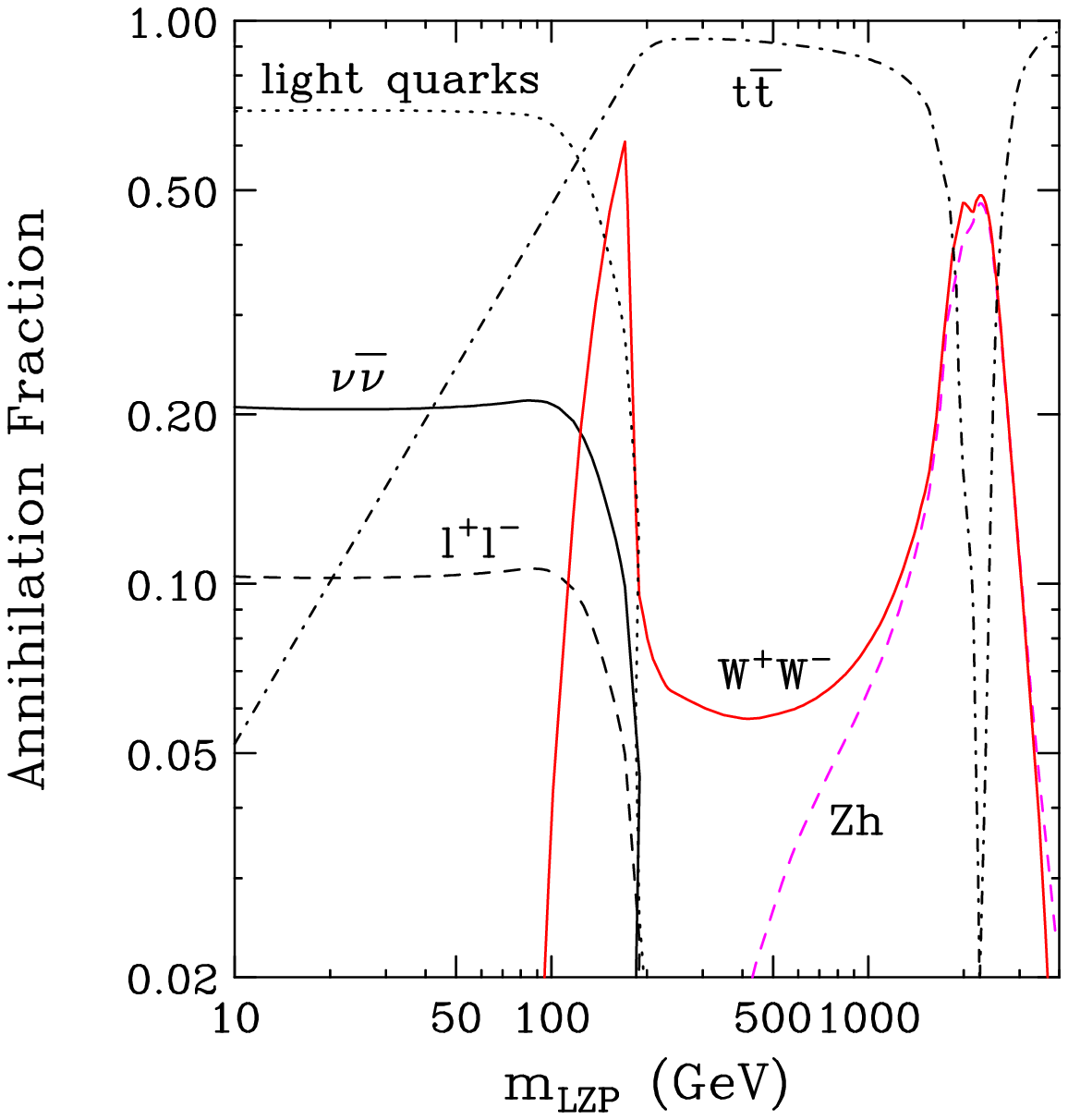}
\hfill
\includegraphics[width=3.2in]{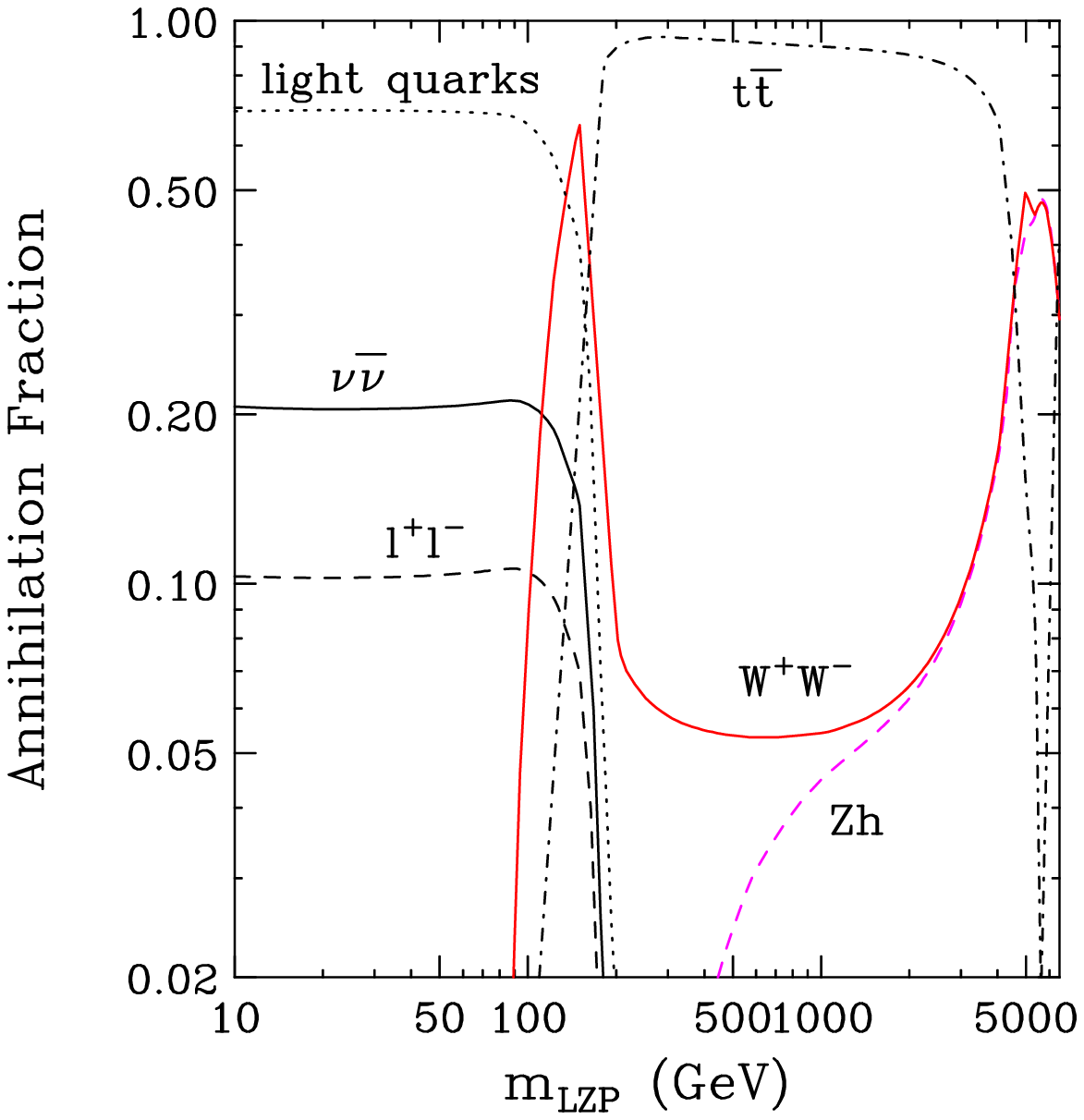}}
\caption{The fraction of LZP annihilations to various final states. Results are shown for $m_{\rm{KK}}=4$ and 10 TeV in the left and right frames, respectively.}
\label{annfrac}
\end{figure}

The next step is to determine the spectrum of neutrinos generated in the annihilation of LZPs. This depends on the region of parameter space considered. In low mass models ($m_{\rm{LZP}} \lsim 100$ GeV), LZPs annihilate primarily through s-channel $Z$ exchange (via $Z -Z^{\prime}$ mixing), and therefore the annihilation modes very nearly follow the decay modes of the $Z$, {\it ie.} 6.7\% to neutrino pairs of each flavor, 3.4\% to charged lepton pairs of each flavor and the remaining $\sim$70\% to quark pairs. For more massive LZPs, annihilations to $t \bar{t}$, $W^+ W^-$ and $Z h$ typically dominate (see figure \ref{annfrac}).

For the first case ($m_{\rm{LZP}} \lsim 100$ GeV), the neutrinos generated will be of sufficiently low energy that one could safely neglect their interactions with the solar medium. Since the experimental (muon) energy thresholds of high-energy neutrino telescopes are in the range of 10 to 100 GeV, few if any of the neutrinos produced in the decays of quarks or charged leptons produced will yield observable muons.\footnote{
Neutrino telescopes detect muons generated in $\nu_{\mu}$ charged current interactions. These muons appear as tracks through the detector volume. Electrons produced by $\nu_e$'s simply generate an electromagnetic shower 
rather than an extended track.  Showers can in principle be observed, but the 
energy thresholds for such events are a few TeV, so are not of interest to 
us here. $\tau$'s decay too quickly to produce observable tracks. Only at PeV 
energies do $\tau$'s travel further than  $\sim$ 100 meters. Again, decaying $\tau$'s can 
produce showers, but well below the energy threshold. All neutrino flavors can produce hadronic showers via neutral current interactions. For our purposes here, only the $\nu_{\tau}$'s and $\nu_e$'s which  have  oscillated into  $\nu_{\mu}$'s can be detected. Very few $\nu_{\mu}$ will come from $\nu_e$ because of the small corresponding mixing angle.} Focusing on annihilations directly to neutrino pairs, the observed muon neutrino flux is a delta function at the mass of the LZP:
\begin{equation}
\frac{dN_{\nu_{\mu}}}{dE_{\nu_{\mu}}} = \frac{2 \Gamma A_{\nu \bar{\nu}} \delta(m_{\rm{LZP}})}{4 \pi R^2_{\rm{ES}}},
\end{equation}
where $\Gamma$ is the rate of LZP annihilations in the Sun, $A_{\nu \bar{\nu}} \approx 0.067$ is the fraction of annihilations to $\nu_{\mu} \bar{\nu}_{\mu}$ (after including neutrino oscillations, $A_{\nu \bar{\nu}}$ is the average of the annihilation fractions to muon and tau neutrinos) and $R_{\rm{ES}}$ is the Earth-Sun distance.

For heavier LZPs, annihilations to top quark pairs often dominate. Top quarks decay nearly 100\% of the time to a $W^{\pm}$ and $b$, each of which can generate neutrinos in their decays.
The neutrino spectrum from this channel is given by \cite{jungman}:
\begin{equation}
\bigg(\frac{dN_{\nu}}{dE_{\nu}}\bigg)_{t \bar{t}} =\frac{2 \Gamma A_{t \bar{t}}}{4 \pi R^2_{\rm{ES}}}\bigg[ \bigg(\frac{dN_{\nu}}{dE_{\nu}}\bigg)_{t
 \rightarrow W^{\pm}} + \,\,\,\, \bigg(\frac{dN_{\nu}}{dE_{\nu}}\bigg)_{t \rightarrow b}\bigg],
\end{equation}
where the first of these terms is given by:
\begin{eqnarray}
\bigg(\frac{dN_{\nu}}{dE_{\nu}}\bigg)_{t \rightarrow W^{\pm}} = \frac{B_{W\rightarrow \
\nu l^{\pm}} \,\, m_t}{(m_t^2 - m_W^2) \gamma \beta } &\times& \ln\bigg(\frac{\rm{min}(E_{\nu
}/(\gamma \,(1-\beta)), m_t/2)}{\rm{max}(E_{\nu}/(\gamma \,(1+\beta)), m_W^2/2 m_t)}\bigg), 
\nonumber \\ 
 &\rm{for}&\,\gamma \,(1-\beta) \, m_W^2/2 m_t \, < E_{\nu} <  \gamma \,(1+\beta) \, m_t/2, 
\end{eqnarray}
where $\gamma=m_{LZP}/m_t$ is the Lorentz boost of the produced top quark and $\beta$ is the 
velocity corresponding to this Lorentz factor. $B_{W\rightarrow \
\nu l^{\pm}}\approx 0.107$ for each neutrino flavor. The second term is given by:
\begin{eqnarray}
\bigg(\frac{dN_{\nu}}{dE_{\nu}}\bigg)_{t \rightarrow b} = \frac{B_{b \rightarrow \nu l^
{\pm} X} \, m_t}{\gamma\, \beta \,0.73 (m_t^2-m_W^2)} &\times& \bigg(3 x^2 - 3 y^2 +4 x^3/3 
-4 y^3/3 + 2 \ln\frac{y}{x} \bigg),  \nonumber \\ 
 &\rm{for}&\, E_{\nu} < 0.73\,\gamma \, (1+\beta) (m_t^2-m_W^2)/2m_t , 
\end{eqnarray}
where $x=2 m_t \,E_{\nu}/(0.73 \,\gamma (1+\beta) (m_t^2+m_W^2))$ and $y=\rm{min}(1, \,2 m_t
 E_{\nu}/(0.73 \,\gamma (1-\beta) (m_t^2+m_W^2)))$. $B_{b \rightarrow \nu l^{\pm} X} \approx 0.103$ for electron and muon flavors and $\sim 0.025$ for tau flavor.

After this spectrum of neutrinos is generated, it must be propagated through the Sun. Above $E_{\nu}\sim 100$ GeV, this can be an important effect. The suppression of the neutrino flux as a result of absorption via charged current interactions in the Sun is given by \cite{crotty}:
\begin{eqnarray}
\frac{d\Phi_{\nu_e}}{dE_{\nu}}(E_{\nu}) \approx \frac{d\Phi^{\rm{ini}}_{\nu_e}}{dE_{\nu}}(E_{\nu}) \times e^{-E_{\nu}/130 \, \rm{GeV}}, \nonumber \\
\frac{d\Phi_{\bar{\nu_e}}}{dE_{\nu}}(E_{\nu}) \approx \frac{d\Phi^{\rm{ini}}_{\bar{\nu_e}}}{dE_{\nu}}(E_{\nu}) \times e^{-E_{\nu}/200 \, \rm{GeV}},  \nonumber \\
\frac{d\Phi_{\nu_{\mu}}}{dE_{\nu}}(E_{\nu}) \approx \frac{d\Phi^{\rm{ini}}_{\nu_{\mu}}}{dE_{\nu}}(E_{\nu}) \times e^{-E_{\nu}/130 \, \rm{GeV}}, \nonumber \\
\frac{d\Phi_{\bar{\nu_{\mu}}}}{dE_{\nu}}(E_{\nu}) \approx \frac{d\Phi^{\rm{ini}}_{\bar{\nu_{\mu}}}}{dE_{\nu}}(E_{\nu}) \times e^{-E_{\nu}/200 \, \rm{GeV}}, 
\end{eqnarray}
where the spectra with the label, {\it ini}, are the initial neutrino spectra injected in the the core of the Sun. 

The effects of propagation on the tau neutrino spectrum are more complicated. Charged current interactions of tau neutrinos produce a tau lepton which, unlike muons or electrons, will decay generating another tau neutrino before coming nearly to rest. As a result of this regeneration process, the total number of tau neutrinos injected near the center of the Sun equals the number that escape, although many of these may have lost much of their energy. The unscattered component of the tau neutrino flux is similar to that of muon and electron neutrinos \cite{crotty}:
\begin{eqnarray}
\frac{d\Phi_{\nu_{\tau}}}{dE_{\nu}}(E_{\nu}) \approx \frac{d\Phi^{\rm{ini}}_{\nu_{\tau}}}{dE_{\nu}}(E_{\nu}) \times e^{-E_{\nu}/160 \, \rm{GeV}}, \nonumber \\
\frac{d\Phi_{\bar{\nu_{\tau}}}}{dE_{\nu}}(E_{\nu}) \approx \frac{d\Phi^{\rm{ini}}_{\bar{\nu_{\tau}}}}{dE_{\nu}}(E_{\nu}) \times e^{-E_{\nu}/230 \, \rm{GeV}}.
\end{eqnarray}
This is accompanied, however, by a spectrum of tau neutrinos which have been scattered in the solar medium. The distribution of the energies of the scattered component of tau neutrinos is roughly described by a log-normal distribution \cite{crotty}:
\begin{equation}
\frac{dN}{dE}=\frac{1}{\sqrt{2 \pi} \ln 10 \, \sigma \,E} \exp\bigg[-\frac{1}{2 \sigma^2} \log^2 \bigg(\frac{E}{E_t}\bigg) \bigg],
\end{equation}
with $E_t=60$ GeV and $\sigma=0.53$ for tau neutrinos and $E_t=113$ GeV and $\sigma=0.49$ for tau anti-neutrinos. 

\begin{figure}[t]
\centering\leavevmode
\mbox{
\includegraphics[width=3.2in]{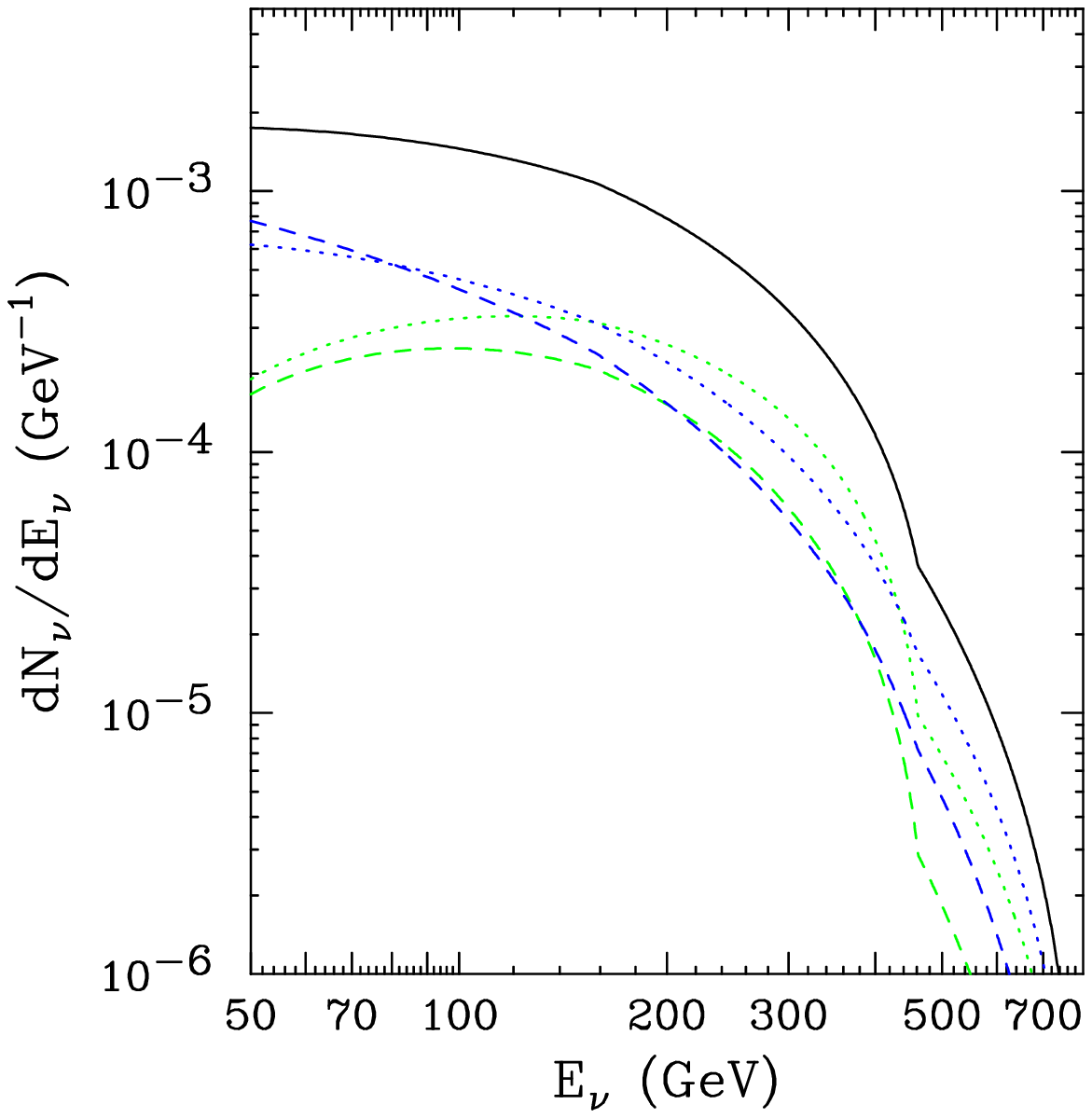}
\hfill
\includegraphics[width=3.2in]{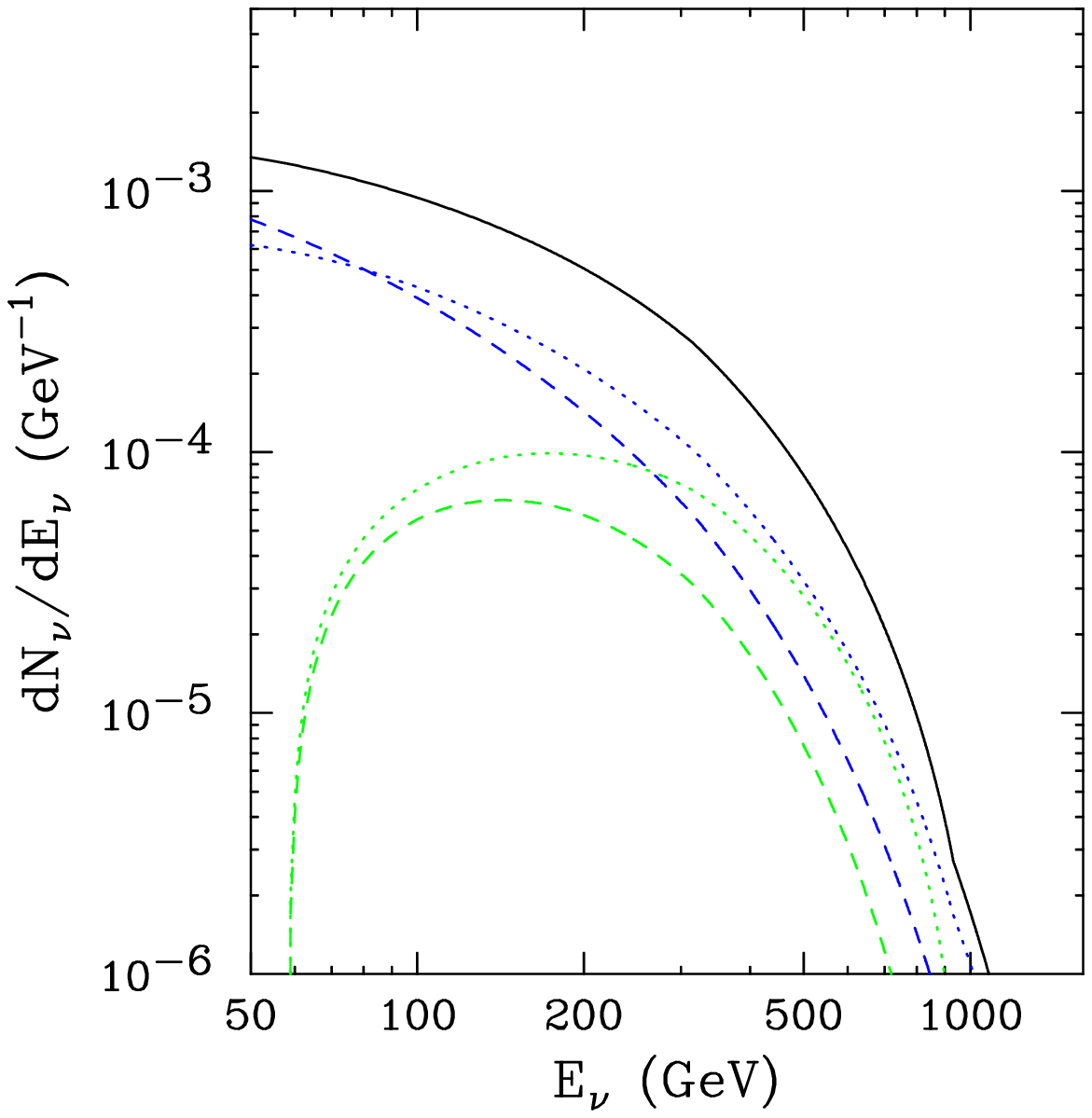}}
\caption{The spectrum of neutrinos and anti-neutrinos per LZP annihilation to top quark pairs in the Sun. The effects of solar propagation have been taken into account. Results in the left and right frames correspond to LZPs with masses of 800 and 1600 GeV, respectively. Light (green) dashed, light (green) dotted, dark (blue) dashed and dark (blue) dotted lines denote $\nu_{\mu}$, $\bar{\nu_{\mu}}$, $\nu_{\tau}$ and $\bar{\nu_{\tau}}$, respectively.  The solid black line is the sum of these four contributions. We have used an annihilation cross section of 3 $\times 10^{-26}$ cm$^3$/s.}
\label{neuspec}
\end{figure}

In figure~\ref{neuspec} we show the final spectrum of neutrinos (plus anti-neutrinos) for LZP annihilations to $t\bar{t}$.

To detect a flux of high-energy neutrinos, large volume neutrino telescopes have been developed \cite{nureview}. Muon neutrinos interacting in, or near, such an experiment generate energetic muons via charged current interactions which emit radiation as they propagate through the detector's Cerenkov medium, such as ice or water. The rate of neutrino-induced muons observed in a high-energy neutrino telescope is found by: 
\begin{equation}
N_{\rm{events}} \simeq \int \int \frac{dN_{\nu_{\mu}}}{dE_{\nu_{\mu}}}\, \frac{d\sigma_{\nu N}}{dy}(E_{\nu_{\mu}},y) \,R_{\mu}((1-y)\times E_{\nu})\, A_{\rm{eff}} \, dE_{\nu_{\mu}} \, dy,
\end{equation}
where $\sigma_{\nu N}(E_{\nu_{\mu}})$ is the neutrino-nucleon charged current interaction cross section \cite{sigmanu}, $(1-y)$ is the fraction of neutrino energy which goes into the muon, $A_{\rm{eff}}$ is the effective area of the detector, $R_{\mu}((1-y)\times E_{\nu})$ is the distance a muon of energy, $(1-y)\,E_{\nu}$, travels before falling below the muon energy threshold of the experiment, called the muon range \cite{range}. 

Currently, the AMANDA telescope at the South Pole has been taking data for a number of years. With a muon energy threshold of approximately 25 GeV, AMANDA may be sensitive to LZPs in the lower mass range, $m_{\rm{LZP}} \approx 35-50$ GeV. Using an effective area of 50,000 square meters for the AMANDA experiment, we estimate an event rate of:
\begin{equation}
N_{\rm{events}} \approx 2.9 \, \rm{yr}^{-1} \,\,\bigg(  \frac{\sigma_{\mathrm{H, SD}}+ \sigma_{\mathrm{H, SI}}
+ 0.067 \, \sigma_{\mathrm{He, SI}} } {10^{-6}\, \mathrm{pb}} \bigg),
\end{equation}
for a LZP with a mass of 40-50 GeV. The ANTARES experiment, currently under construction in the Mediterranean, is designed to have a somewhat lower energy threshold ($\sim$10 GeV) and an 100,000 square meter effective area. For ANTARES, we estimate an event rate of:
\begin{equation}
N_{\rm{events}} \approx 10.5 \, \rm{yr}^{-1} \,\,\bigg(  \frac{\sigma_{\mathrm{H, SD}}+ \sigma_{\mathrm{H, SI}}
+ 0.067 \, \sigma_{\mathrm{He, SI}} } {10^{-6}\, \mathrm{pb}} \bigg), 
\end{equation}
for a LZP with a mass of 40-50 GeV.

\begin{figure}[t]
\centering\leavevmode
\includegraphics[width=3.5in]{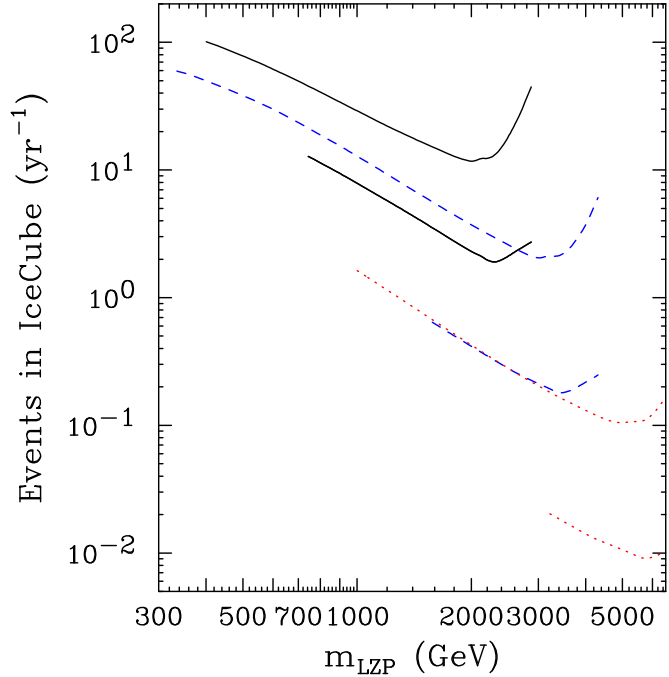}
\caption{The number of events per year in a detector with a cubic kilometer effective area, such as IceCube. A 50 GeV muon energy threshold has been used. Solid (black) lines, dashed (blue) lines and dotted (red) lines correspond to $M_{\rm{KK}}=$4, 6 and 10 TeV, respectively. For each value of $M_{\rm{KK}}$, two lines are shown. These upper and lower lines use parameters corresponding to a `median' and `minimum' set of couplings (see Table \ref{datasets-table} for more details). Values are shown only over the mass ranges which do not overproduce the density of dark matter.}
\label{icecubeevents}
\end{figure}

For heavier LZPs, which annihilate dominantly to $t \bar{t}$, $W^+ W^-$ or $Zh$, larger volume neutrino telescopes will most likely be needed. The IceCube experiment, currently under construction at the South Pole, will have a full cubic kilometer of instrumented volume. The event rate predicted in IceCube are shown in figure~\ref{icecubeevents}.

These rates must compete with a background of atmospheric neutrinos \cite{atmback} and neutrinos generated in cosmic ray interactions in the Sun's corona \cite{sunback}. After cuts, approximately 5 to 10 atmospheric neutrino events are expected per year in AMANDA within the solid angle in the direction of the Sun. The background rate from cosmic ray interactions is expected to be significantly lower. The current data collected by AMANDA should be sufficient to exclude $\sim$50 GeV LZPs with elastic scattering cross sections larger than $\sim 10^{-6}$ pb, although a proper analysis of the data would need to be done to calculate a precise limit.

The background rate of atmospheric neutrinos in ANTARES will be somewhat higher than the rate in AMANDA due to their larger area and lower energy threshold. When completed, ANTARES should be slightly more sensitive to $\sim$50 GeV LZPs than AMANDA.

IceCube, with a background rate of $\sim$100 atmospheric neutrino events per year, will be able to test models which predict more than $\sim$10 events per year (see figure~\ref{icecubeevents}). With several years of exposure, perhaps a rate of a few events per year could be detected.

\section{Signatures of LZP Annihilations in Cosmic Positrons}
\label{sec:positrons}

High-energy positrons can be produced in LZP annihilations throughout the Galactic halo. These positrons then travel under the influence of tangled Galactic magnetic fields, losing energy through inverse Compton scattering and synchrotron processes. The positron spectrum observed can be arrived at by solving the diffusion-loss equation:
\begin{equation}
\frac{\partial}{\partial t}\frac{dn_{e^{+}}}{dE_{e^{+}}} = \vec{\bigtriangledown} \cdot \bigg[K(E_{e^{+}},\vec{x})  \vec{\bigtriangledown} \frac{dn_{e^{+}}}{dE_{e^{+}}} \bigg]
+ \frac{\partial}{\partial E_{e^{+}}} \bigg[b(E_{e^{+}},\vec{x})\frac{dn_{e^{+}}}{dE_{e^{+}}}  \bigg] + Q(E_{e^{+}},\vec{x}),
\label{dif}
\end{equation}
where $dn_{e^{+}}/dE_{e^{+}}$ is the number density of positrons per unit energy, $K(E_{e^{+}},\vec{x})$ is the diffusion constant, $b(E_{e^{+}},\vec{x})$ is the rate of energy loss and $Q(E_{e^{+}},\vec{x})$ is the spatial and energy distribution of positrons injected through LZP annihilations.

We parameterize the diffusion constant \cite{diffusion,posbaltz} by:
\begin{equation}
K(E_{e^{+}}) = 3.3 \times 10^{28} \bigg[3^{0.47} + E_{e^{+}}^{0.47} \bigg] \,\rm{cm}^2 \, \rm{s}^{-1},
\label{k}
\end{equation}

and the energy loss rate by:
\begin{equation}
b(E_{e^{+}}) = 10^{-16} E_{e^{+}}^2 \,\, \rm{s}^{-1}.
\label{b}
\end{equation}
$b(E_{e^{+}})$ is the result of inverse Compton scattering on both starlight and the cosmic microwave background \cite{lossrate}. We impose the boundary conditions for our diffusion zone corresponding to a slab of thickness $2L$, where $L$ is fit to observations to be approximately 4 kpc \cite{diffusion}. Beyond the boundaries of our diffusion zone, we drop the positron density to zero (free escape boundary conditions). These diffusion parameters are constrained by analysis of stable nuclei in cosmic rays (primarily by fitting the boron to carbon ratio) along with radioactive secondaries in the cosmic ray spectrum \cite{ptuskin}.

In addition to the propagation effects described above, as positrons approach the solar system, their interaction with the solar wind and magnetosphere can be important. These effects, called solar modulation, may considerably effect the positron spectrum at energies below 10-20 GeV.   

The spectrum of positrons injected into the halo depends on the dominant annihilation modes of the LZP. For somewhat light LZPs ($m_{\rm{LZP}} \lsim 100$ GeV), annihilations produce positrons directly ($e^+ e^-$) as well as through the decays of muon, tau and quark pairs. The annihilations to $e^+ e^-$, and to a lesser extent $\mu^+ \mu^-$ and $\tau^+ \tau^-$, generate a very hard spectrum of positrons. Quark decays typically generate positrons with lower energies. For heavier LZPs, which annihilate to $t \bar{t}$, $W^+ W^-$ or $Zh$, a softer and less distinctive positron spectrum is generated. We have used PYTHIA \cite{pythia}, as implemented in DarkSusy program \cite{darksusy}, to generate the injected positron spectra.

Finally, the spatial distribution of positrons injected into the halo depends on the halo model assumed and on the quantity of dark matter substructures distributed in the local halo. Since positrons in the range of energies considered here typically lose most of their energy over distances greater than a few kiloparsecs, this channel does not efficiently sample annihilations near the Galactic center. For this reason, the presence of a cusp in the Galactic halo distribution is of only secondary importance. We have used a standard Navarro-Frenk-White halo profile in our calculations \cite{nfw}.

The presence of dark substructure in the local halo can lead to enhancements in the dark matter annihilation rate. Analytic estimates of this effect find that the positron flux is likely to be boosted by a factor of 2 to 5 due to such effects \cite{berezinsky}. Recently, however, it has been suggested that greater quantities of substructure may have formed \cite{mooreclumps} and, depending on how many of these object survive \cite{zhao}, considerably larger boost factors may be possible.

The HEAT experiment (High-Energy Antimatter Telescope) has made the most detailed measurements of the high-energy cosmic positron spectrum to date. This data appears to contain an excess in the spectrum, peaking at around 7-10 GeV and extending to higher energies \cite{heat}.  Although the source of these positrons is not known, it has been often suggested that this signal could be the product of dark matter annihilations \cite{positrons,positrons2,positrons3,posbaltz,Hooper:2004xn}.

\begin{figure}[t]
\centering\leavevmode
\includegraphics[width=3.5in]{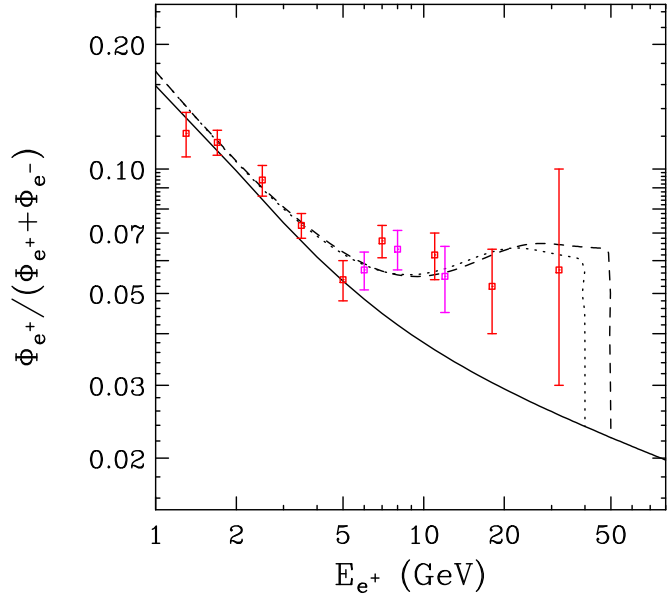}
\caption{The cosmic positron fraction from LZP annihilations for masses of 40 (dots) and 50 (dashes) GeV. In each case, the normalization of the dark matter contribution was chosen to maximize the quality of the fit to the HEAT data (shown as error bars). We find $\chi^2$'s of approximately 1.1 per degree of freedom for each case. Using an annihilation cross section of $3 \times 10^{-26}$ cm$^3$/s, and a local dark matter density of $\rho$(local)=0.4 GeV/cm$^3$, boost factors of 23 and 17 were needed to fit the data. The solid line is the background-only prediction.}
\label{posheatfig}
\end{figure}

In figure~\ref{posheatfig}, we show the positron spectrum from LZP annihilations compared to the HEAT data. The results are shown as the ratio of positrons to positrons plus electrons, called the positron fraction. Included are the primary and secondary electron fluxes as well as the secondary positron flux calculated in the model of Ref.~\cite{secbg}. The solid line in the figure is for this background alone, with no dark matter component. The dashed and dotted lines correspond to LZP masses of 50 and 40 GeV, respectively. These have been normalized using boost factors of 23 and 17, respectively, assuming a local dark matter density of $\rho = 0.4$ GeV/cm$^3$. These models both fit the HEAT data very well, yielding $\chi^2 \approx 1.1$ per degree of freedom. 

Although considerably heavier LZPs cannot produce the excess observed by HEAT without unnaturally large boost factors, these models may still be accessible to future cosmic positron experiments, which will be considerably more sensitive to dark matter annihilations \cite{futureposexpt}. 

The PAMELA experiment, to be launched into orbit later this year, is a satellite borne experiment designed to study the matter-antimatter asymmetry of the universe with very precise cosmic ray measurements. PAMELA's primary objectives include the measurement of the cosmic positron spectrum up to 270 GeV \cite{pamela}, well beyond the range studied by HEAT. This improvement is made possible by PAMELA's large acceptance (20.5 cm$^2$\,sr \cite{picozza}) and long exposure time (3 years).

The AMS-02 experiment has been designed to be deployed on the International Space Station (ISS) for a three year mission sometime around the end of the decade. AMS-02's acceptance of about 450 cm$^2$\,sr is considerably larger than PAMELA's. This, along with superior energy resolution and electron and anti-proton rejection, makes AMS-02 the premiere experiment for measuring the cosmic positron spectrum.

\begin{figure}[t]
\centering\leavevmode
\mbox{
\includegraphics[width=3.2in]{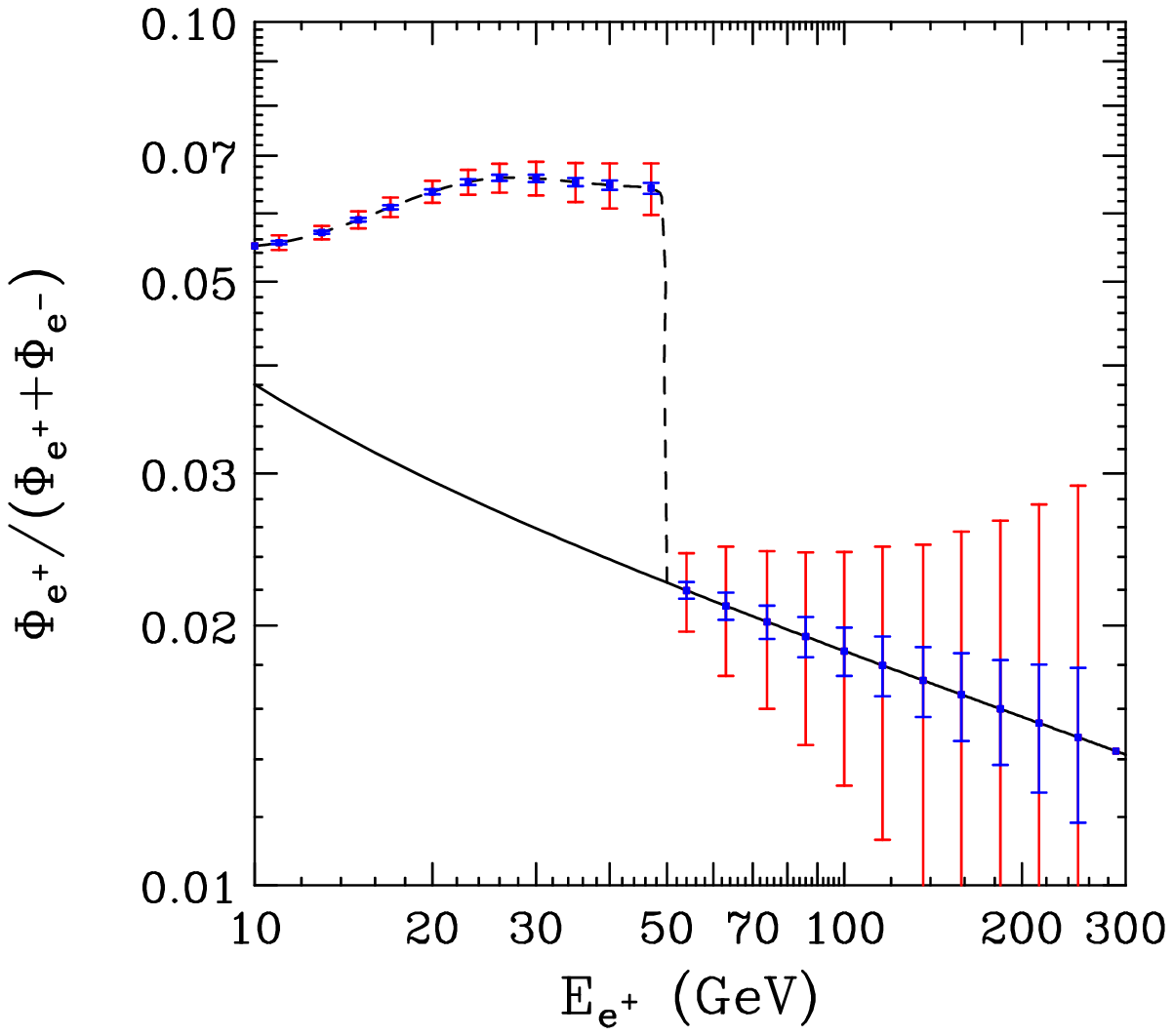}
\hfill
\includegraphics[width=3.2in]{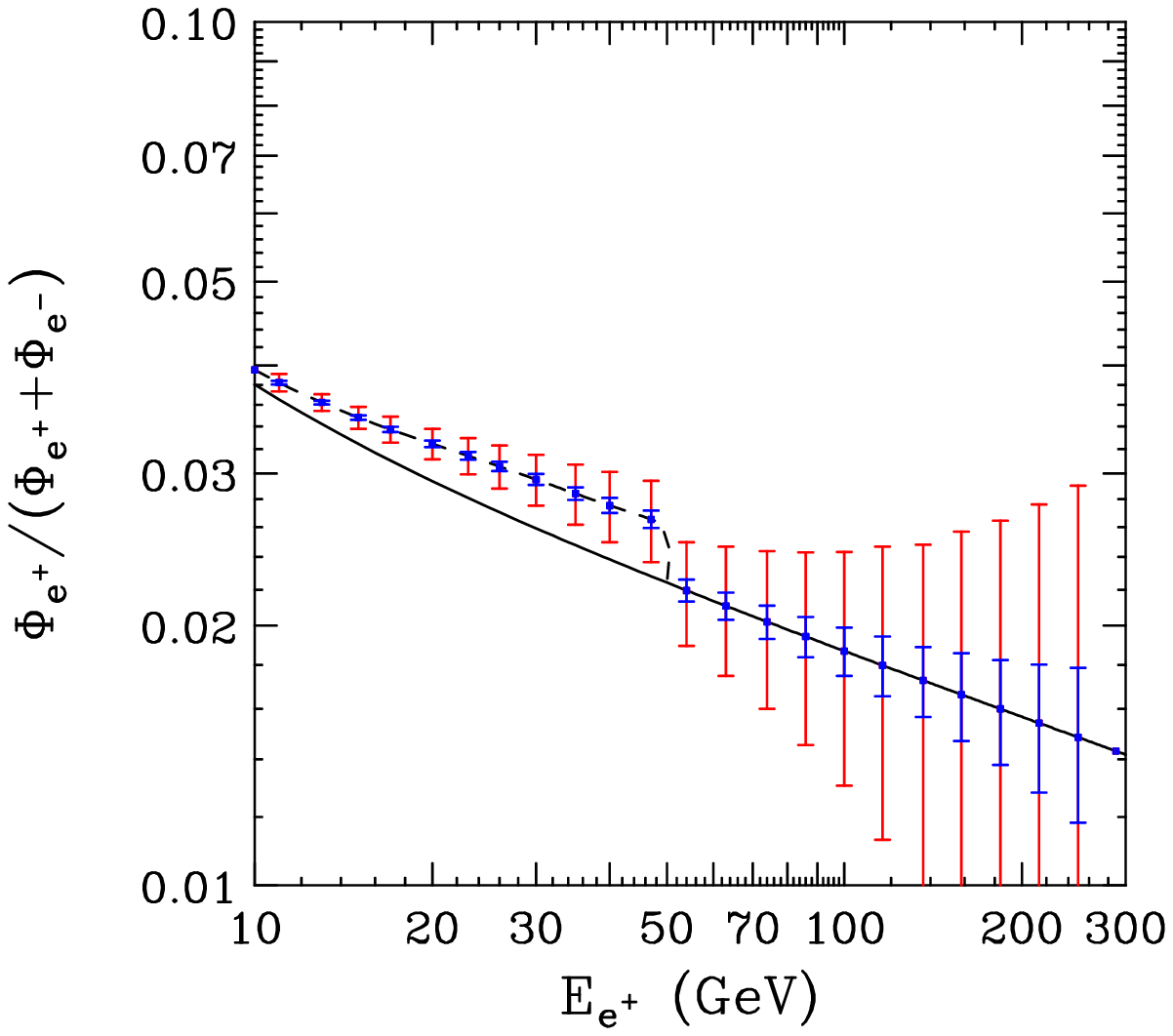}}
\caption{The cosmic positron fraction from 50 GeV LZP annihilations compared to the error bars projected for the PAMELA (larger, red) and AMS-02 (smaller, blue) experiments. In the left frame, the spectrum is normalized to the HEAT data (see figure~\ref{posheatfig}) with a boost factor of 23. In the right frame, a conservative boost factor of 2 was used. The solid line is the background-only prediction.
}
\label{posfuture50}
\end{figure}

If LZP annihilations are in fact responsible for the HEAT excess, PAMELA and AMS-02 should be able to unambiguously identify this fact. In the left frame of figure~\ref{posfuture50}, we show the spectrum of figure~\ref{posheatfig}, with the error bars projected for the PAMELA (larger, red) and AMS-02 (smaller, blue) experiments. In addition to the overall precision of these measurements, the distinctive feature at the LZP mass would be unmistakable to these experiments. In the right frame of figure~\ref{posfuture50}, we show this same model, but with a conservative boost factor of 2. Even in this case, the distinctive spectral feature at the LZP mass can be confidently identified. It may also be possible to identify a discrete feature in the cosmic positron spectrum using Atmospheric Cerenkov Telescopes (ACTs)~\cite{Baltz:2004ie}.

\begin{figure}[t]
\centering\leavevmode
\includegraphics[width=3.5in]{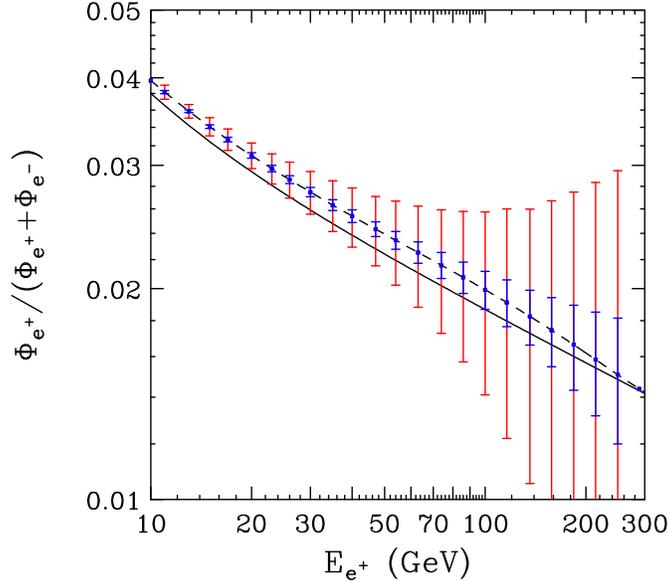}
\caption{The cosmic positron fraction from LZP annihilations for a mass of 400 GeV annihilating to top quark pairs compared to the error bars projected for the PAMELA (larger, red) and AMS-02 (smaller, blue) experiments. We have used an annihilation cross section of $3 \times 10^{-26}$ cm$^3$/s, a local dark matter density of $\rho$(local)=0.4 GeV/cm$^3$ and a boost factor of 10.  The solid line is the background-only prediction.}
\label{pos400}
\end{figure}

If LZPs are heavier and annihilate mostly to top quarks, or gauge and higgs bosons, positrons will be produced in the subsequent decay chains, {\it i.e.} $t \rightarrow W^{\pm} b$, $W^+ \rightarrow e^+ \nu$, $b \rightarrow e^+ \nu X$, etc. The resulting spectrum, for a 400 GeV LZP annihilating to $t \bar{t}$, is shown in figure~\ref{pos400}. Although no highly distinctive features are present in this spectrum, it may still be possible to statistically identify such a spectrum over the background prediction.

Other anti-matter channels for indirect dark matter searches have been studied as well, namely cosmic anti-proton and anti-deuteron studies. Although such searches are indeed interesting, they sample over considerably larger regions of the Galactic halo, making the choice of diffusion zone boundary conditions and halo profile models very important. These ambiguities make predictions of anti-proton and anti-deuteron spectra from dark matter annihilations difficult to reliably calculate.

\section{Gamma-Rays From LZP Annihilations Near the Galactic Center}
\label{sec:GC}

LZP annihilations can produce high-energy gamma-rays through the hadronization and decay of pions generated in the cascading of their annihilation products. In addition to this continuum emission, LZP annihilations can also produce $\gamma \gamma$ and $\gamma Z$ final states, for example from box diagrams with a $X_s $ along one side and $t$ along the other three. Such processes would yield the very distinctive feature of mono-energetic gamma-ray lines, although with a much smaller cross section than for continuum emission. The full cross sections for these processes have not yet been calculated for this model. For these reasons, we will focus on continuum gamma-rays in this section.

Unlike positrons and other charged particles, cosmic gamma-rays travel in straight lines and can travel greater distances without energy loss. For these reasons, regions of the Galaxy which are expected to have higher concentrations of dark matter (and higher annihilation rates) can be studied with gamma-ray telescopes without integrating over the entire Galactic volume. In particular, the Galactic center is a promising region for such observations.

The
$\gamma$-ray flux from dark matter annihilations near the Galactic
center is given by
\begin{equation}
\Phi_\gamma (\psi, E_\gamma) = \langle\sigma v\rangle
                               \frac{\rm{d}N_\gamma}{\rm{d} E_\gamma}
                               \frac{1}{4\pi m_{\rm{LZP}}^2}\int_{\rm los}
                               \rm{d}l(\psi)\ \rho^2(r).
\end{equation}
Here, $\psi$ is the angle between the line-of-sight (los) and the
Galactic center, $\langle\sigma v\rangle$ is the LZP
annihilation cross-section and $\rho(r)$ is the dark matter density at distance, $r$, from the
Galactic center. $dN_{\gamma}/dE_{\gamma}$ is the spectrum of continuum gamma-rays produced in each annihilation.

This expression can be further separated into
two factors, one specifying the particle physics model (mass,
cross-section and fragmentation spectrum) and another describing
the dark matter distribution. Normalizing to the distance to the
Galactic center and the local halo dark matter density, the latter
factor can be written as:
\begin{equation}
 J (\psi) = \frac{1}{8.5\,\rm{kpc}}
            \left(\frac{1}{0.3\,\rm{GeV/cm}^3}\right)^2 \int_{\rm los}
            \rm{d}l(\psi)\ \rho^2(l).
\end{equation}

Defining $\overline{J(\Delta \Omega)}$ as the average of
$J(\psi)$ over the solid angle $\Delta \Omega$ (centered on $\psi=0$),
we can write
\begin{equation}
 \Phi_\gamma (\psi, E_\gamma) \simeq 5.6 \times 10^{-12}\
 \rm{cm}^{-2}\rm{s}^{-1} \frac{\rm{d}N_\gamma}{\rm{d}E_\gamma}
 \left(\frac{\langle\sigma
 v\rangle}{3\times10^{-26}\rm{cm}^3\rm{s}^{-1}}\right)
 \left(\frac {1\,\rm{TeV}}{m_{\rm{LZP}}}\right)^2 \overline{J (\Delta\Omega)}
 \Delta\Omega.
\end{equation}
For a standard Navarro-Frenk-White (NFW) halo profile, averaged over a solid angle of $\Delta\Omega \sim 5\times10^{-5}$ sr, we calculate a value of $\overline{J(5 \times 10^{-5}\, \rm{sr})} \simeq 5.6 \times 10^3$. This can vary considerably with halo model, however. For example, with a Moore {\it et al} profile \cite{moore}, we calculate a much larger value of $\overline{J(5 \times 10^{-5}\, \rm{sr})} \simeq 1.9 \times 10^6$. If the effects of adiabatic compression or accretion onto the central supermassive black hole are taken into account, even larger values can be found \cite{largerJ}. 

\begin{figure}[t]
\centering\leavevmode
\includegraphics[width=3.5in]{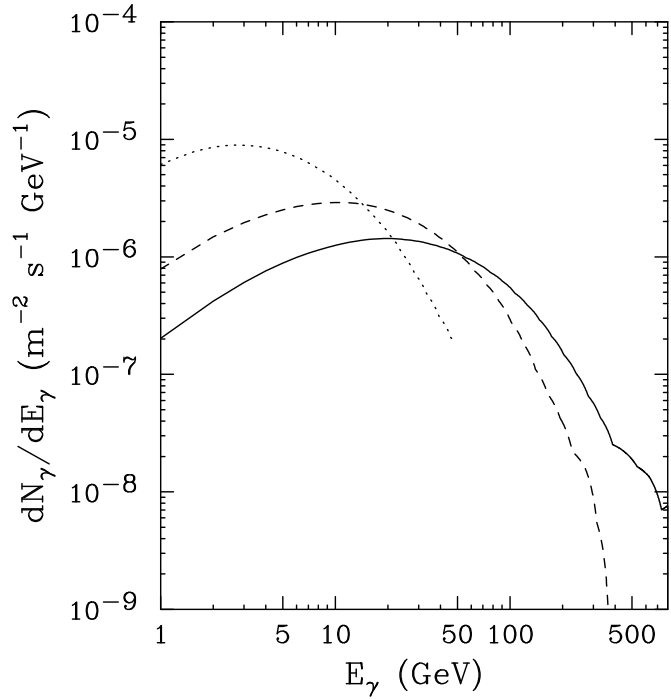}
\caption{The gamma-ray spectrum from LZP annihilations in the Galactic center region. The dotted, dashed and solid lines are for LZP masses of 50, 400 and 800 GeV. Annihilations over an angular cone of size $\Delta\Omega \sim 5\times10^{-5}$ sr were included. A Navarro-Frenk-White halo profile has been use as well as  an annihilation cross section of 3 $\times 10^{-26}$ cm$^3$/s.}
\label{gammaspec}
\end{figure}

The spectrum of gamma-rays predicted from LZP annihilations is shown in figure~\ref{gammaspec}. This result is not substantially different than what would be expected from the annihilation of neutralinos or other WIMPs. Also, recent observations of a gamma-ray spectrum extending above $\sim$10 TeV may pose serious challenges to identifying any potential gamma-ray signal from WIMP annihilations \cite{hess}, assuming that this is not in fact the annihilation signature of a very heavy WIMP \cite{heavyhess}. For these reasons, we do not further discuss the gamma-ray indirect detection channel here.

\section{Comparisons to SUSY and UED Models}

In this section, we briefly summarize the phenomenological features relevant to dark matter detection for an LZP Dirac RH neutrino and compare these features to those of the lightest neutralino in supersymmetric models and the first Kaluza-Klein excitation of the hypercharge gauge boson in models with Universal Extra Dimensions.

\vspace{0.5cm}
{\it Mass Range}: LZPs can generate the observed quantity of dark matter thermally in two mass ranges: near the $Z$-resonance ($m_{\rm{LZP}} \approx$ 35-50 GeV) and for considerably heavier masses ($m_{\rm{LZP}} \gsim$ several hundred GeV) \cite{Agashe:2004ci,Agashe:2004bm}. Neutralinos can span a very wide range of masses, although in the models most often studied, they fall in the range of 50-1000 GeV. The LKP in UED models requires a mass in the range of 700-900 GeV in order to generate the observed thermal relic density, unless other KK modes participate in the freeze-out process \cite{Servant:2002aq}. If this is the case, somewhat smaller masses are possible. A  recent analysis, taking into account the effects of second level KK modes, indicates that the upper edge of this mass range is favored \cite{Kakizaki:2005en}. In any case, we should keep in mind that the precise prediction of the LKP relic density depends on the particular KK mass spectrum which is used and is somewhat model-dependent.

\vspace{0.5cm}
{\it Annihilation Cross Section}: If any of these dark matter candidates generate the observed relic density thermally, they must have an annihilation cross section of $\langle \sigma v \rangle \approx 3 \times 10^{-26}$ cm$^3$/s evaluated at the freeze-out temperature. For the purpose of indirect detection, we are interested in the annihilation cross section in the low velocity limit, which does not necessarily equal the value at freeze-out, however. For most of the parameter space in this model, these two cross sections are not considerably different. In UED models, they are always the same. For neutralinos, this issue depends on the composition of the LSP and the spectrum of other supersymmetric particles. For the case of neutralino annihilation through s-channel pseudoscalar higgs exchange to fermion pairs, as if often the dominant channel for bino-like LSPs, the cross section at freeze-out will be very similar to the value at low velocities.

\vspace{0.5cm}
{\it Dominant Annihilation Channels}: The leading annihilation channels for LZPs are shown in Fig.~\ref{annfrac}. Below 100 GeV, annihilations to light quarks, neutrinos and charged leptons are important. At higher masses, $t \bar{t}$, $W^+ W^-$ and $hZ$ channels dominate. In UED models, about 20\% of LKP annihilations go to charged lepton pairs of each family. About 1.2\% go to each flavor of neutrino pairs. The remaining annihilations go to quarks and higgs bosons. 

For neutralinos, again the composition and SUSY spectrum are important in predicting the dominant annihilation modes. For a bino-like LSP, annihilations to $b \bar{b}$, $\tau^+ \tau^-$ and, if kinematically allowed, $t \bar{t}$ usually dominate. For winos and higgsinos, annihilations to gauge and higgs bosons are often important. Since neutralinos are Majorana fermions, their annihilations to light fermions are chirality suppressed, unlike for the LZP and LKP.

\vspace{0.5cm}
{\it Elastic Scattering With Nucleons}: LZP capture in the Sun can be dominated by spin-dependent scattering with hydrogen nuclei (protons) or by scalar scattering with helium nuclei. The size of these cross sections can vary a great deal, depending on the choices of parameters. In particular, the scale $M_{\rm{KK}}$ (mass of KK gauge bosons) plays a very important role. The bounds on the spin-independent cross section placed by the CDMS experiment \cite{cdms} are often exceeded for smaller values of $M_{\rm{KK}}\sim$ 3--4 TeV \cite{Agashe:2004ci,Agashe:2004bm}. On the other hand, values as small as $\sim 10^{-9}$ pb are possible for $M_{\rm{KK}}\sim$10 TeV.

For neutralinos, these cross sections can vary from in excess of the CDMS bound, to as small as $\sim 10^{-15}$ pb, although in constrained models, they often fall in the range of  $\sim 10^{-6}$--$10^{-9}$ pb. In UED models, spin-dependent scattering dominates LKP capture in the Sun, with cross sections of $10^{-5}$--$10^{-6}$ pb.

\section{Discussion and Conclusions}

In this article, we have studied the prospects for the detection of a stable, right-handed, Dirac neutrino which can emerge in grand unified models in a Randall-Sundrum spacetime. After introducing into the model a $Z_3$ symmetry necessary to ensure proton stability, the Lightest $Z_3$ charged Particle (LZP) is stable and a viable dark matter candidate. We have studied the indirect detection prospects for such a particle through three channels: neutrinos produced through annihilations in the Sun, positrons produced through annihilations in the Galactic halo and gamma-rays produced through annihilations in the Galactic center.

The prospects for detecting high-energy neutrinos from the Sun in this model are very encouraging. For a light LZP in this model ($m_{\rm{LZP}} \lsim 100$ GeV), annihilations often produce neutrinos directly. The currently operating AMANDA experiment may already be sensitive to some of these models. In the future, the ANTARES experiment will further probe into this region. For heavier LZPs, annihilations to top quark pairs produce neutrinos efficiently in their decays. The kilometer-scale experiment IceCube, currently under construction, should be sensitive to most of these models in which the Kaluza-Klein scale, $M_{\rm{KK}}$, is lower than $\sim6$ TeV (a favorite range as far as fine-tuning in the electroweak sector is concerned).

Lighter LZPs also generate positrons very efficiently. A 35--50 GeV LZP annihilating in the galactic halo produces a spectrum of positrons which matches the spectrum observed by the HEAT experiment quite well, and with more moderate annihilation rates than for neutralino dark matter. At an energy equal to the LZP mass, a discrete feature appears in the positron spectrum which, if identified in future experiments such as PAMELA and AMS-02, would provide an unambiguous signature of dark matter annihilations. 

Finally, LZP annihilations near the Galactic center may provide an observable flux of gamma-rays. We have calculated the spectrum of gamma-rays in this model and find that is it not considerably different than for the case of annihilating neutralinos.

These results could be generalized to other Dirac RH neutrino dark matter particles, though it is not clear in which other contexts such a particle would naturally arise. In 4D, this would require the addition of vector-like neutrinos very weakly coupled to SM gauge bosons but interacting via an extra $U(1)$. The coupling of SM fermions to this extra $U(1)$ should be suppressed to keep their coupling to the $Z$ (due to $Z$-$Z^{\prime}$ mixing) almost unshifted. It is also unclear what would make this particle stable in this context. 
Note, however, that via the AdS/CFT correspondence, our Randall-Sundrum scenario is dual to a 4D composite Higgs scenario, 
in which the unification of gauge couplings has recently been studied  \cite{Agashe:2005vg}. In this case, the LZP maps to some low-lying hadron at the composite scale. We also point out that in Refs.~\cite{Agashe:2004ci,Agashe:2004bm}, the strong coupling scale is close to the curvature scale so that ${\cal O} (1)$ variations in our calculations are expected. These results should therefore be considered as representative rather than a complete description.

Annihilations can vary from one Dirac RH neutrino dark matter model to another, depending on whether, at large $\nu^{\prime}_R$ mass, annihilations take place via s-channel $Z^{\prime}$ exchange only or also via a t-channel $X_s$--type gauge boson. On the other hand, the elastic scattering cross section is mainly model-independent (determined by the $\nu^{\prime}_R -Z$ coupling).

\section*{Acknowledgments}
This work benefitted from earlier collaborations with K.~Agashe. DH is supported by the Leverhulme trust.


\end{document}